\documentclass{aip-cp}

\usepackage[numbers]{natbib}
\usepackage{rotating}
\usepackage{graphicx}

\usepackage{savesym}
\savesymbol{iint}
\savesymbol{iiint}
\savesymbol{iiiint}
\savesymbol{idotsint}
\usepackage{amsmath}
\restoresymbol{AMS}{iint}
\restoresymbol{AMS}{iiint}
\restoresymbol{AMS}{iiiint}
\restoresymbol{AMS}{idotsint}

\newcommand{\ep}{\varepsilon _{\rm 0}}
\newcommand{\omc}{\omega _{\rm c}}

\newcommand{\omps}{\omega _{\rm p}^2}
\newcommand{\kp}{k_{//}}
\newcommand{\vT}{v_{\rm T}}

\newcommand{\abs}[1]{\left\vert#1\right\vert}
\newcommand{\K}{\mathcal{K}}
\newcommand{\Ell}{\mathcal{L}}

\newcommand{\W}{\mathcal{W}}
\newcommand{\V}{\mathcal{V}}

\newcommand{\kpi}{\kappa_\pi} % Elementary poloidal contribution to k_//
\newcommand{\khat}{\breve{\kappa}} % represents kappa modulo 1
\newcommand{\rmd}{{\rm d}}
\newcommand{\rme}{{\rm e}}
\newcommand{\rmi}{{\rm i}}

\begin{document}

% Title portion
\title{Dielectric Kernels for Maxwellian Tokamak Plasmas}

\author{P. U. Lamalle}

\affil{F-04100 Manosque, France}
\corresp{phlama774@gmail.com}

\maketitle

%---------------------------------------------------------------------
\begin{abstract}
New integral kernels describing the full-wave dielectric response of Maxwellian tokamak plasmas are presented. They realistically account for the rotational transform and for wave dispersion in presence of equilibrium magnetic field parallel gradients. These kernels rely on special functions of three variables that generalize the standard plasma dispersion function; their main analytical properties are given, leading to efficient evaluation.
This approach is free from the poloidal Fourier mode expansion of the HF fields which appears in earlier formulations and gives complete freedom for the numerical resolution of the wave equation: it will typically be applied to 2D finite element discretizations, allowing local mesh refinements as required near cyclotron resonance layers and in regions of rapid HF field variations. This first presentation is to lowest order in the Larmor radius for the sake of clarity but will readily generalize to all orders in ($\rho_{LT}/\lambda_\perp$).

\end{abstract}
%---------------------------------------------------------------------

\section{INTRODUCTION}

A number of works addressing wave propagation and absorption in tokamaks, see e.g. \cite{Gambier1985}, \cite{Brambilla1988}, \cite{Lamalle1997} and the derived full-wave codes, are based on toroidal and poloidal Fourier expansions of the HF fields, since this allows accurate modelling of the plasma dielectric properties in presence of rotational transform. A significant drawback of the poloidal expansion is its lack of flexibility, in that it does not allow local refinements of the numerical discretization on a given magnetic surface. As a remedy to this, the present communication reports on new theoretical expressions of the Maxwellian dielectric response which are valid irrespective of the HF field representation in a tokamak meridian plane. The paper lists the main results of this investigation to lowest order in the Larmor radius. A detailed account of the derivation and the generalization to all orders in FLR will be presented elsewhere \cite{Lamalle2019a}.
This work is based on the Galerkin formulation of the plasma wave equation inside the tokamak chamber \cite{Lamalle1997}:
\begin{equation}\label{Galer}
\frac{\rmi}{2} \, \int_\V \left[\, \frac{1}{\omega\,\mu_0} (\nabla\times\boldsymbol{F})^*.(\nabla\times\boldsymbol{E})
-\omega\,\ep\,\boldsymbol{F}^*.\boldsymbol{E} \,\right] \, \rmd r^3 
+ \sum_{\beta} \W_{\boldsymbol{FE}\beta} 
= -\frac{1}{2} \int_\V \boldsymbol{F}^*.\boldsymbol{j}_S  \, \rmd r^3
\end{equation}
in which $\boldsymbol{E}$ is the unknown HF electric field and $\boldsymbol{F}$ an arbitrary test function. With the assumed axisymmetry this expression splits into independent contributions for the different RF field toroidal Fourier modes $\propto \exp{\rmi n \varphi}$, and index $n$ is kept implicit wherever unambiguous. In the poloidal Fourier representation the dielectric response of particle species $\beta$ for mode $n$ is
\begin{equation}\label{CTM323}
\W_{\boldsymbol{FE}\beta}^{(n)}
= \sum\limits_{m_1,\,m_2 = - \infty}^{+\infty} \;\; \sum\limits_{p \, = - \infty}^{ + \infty} \;\; \W_{21\beta}^{p}
\end{equation}
in which $p$ is the cyclotron harmonic index and $m_1$, $m_2$ poloidal mode indices of HF field and test function, respectively. The terms of lowest order in the Larmor radius are only present for $p=\pm 1$ (fundamental cyclotron interactions) and $p=0$ (Landau-\v{C}erenkov interaction). Within standard assumptions discussed in \cite{Lamalle1998} they are given by poloidal Fourier transforms of plasma dispersion functions:
\begin{eqnarray}\label{CTM323bis}
{\W_{21\beta}^{p}} = -\rmi \pi \ep  \, \sum_{L=-1}^{1} \delta_{L,p} \, 2^{\alpha/2}
\int \, \rmd \rho \; 
F_{\Ell,m_2}^\ast \,
\left\{\int \frac{RJ \,\omps}{\abs{\kp}\vT} \; 
I_{\alpha}\left(\frac{\omega - p \omc}{\abs{\kp}\vT}\right) \, 
\rme^{\rmi(m_1-m_2)\theta} \, \rmd\theta \, \right\} \, 
E_{\Ell,m_1 } 
\qquad  (\alpha=2 \delta_{L,0})
\end{eqnarray}
 We use a condensed notation in which the field components $\Ell$ (left-hand, right-hand circular polarizations and parallel component) are `encoded' by the integer $L$: $\, \Ell = +$, $-$, $//$ for $L=+1$, $-1$, $0$, respectively. $I_0(z)=Z_{FC}(z)$ is the standard plasma dispersion function, $I_1(z)=1+z\, I_0(z)$ and $I_2(z)=z \, I_1(z)$.
$RJ$ is the Jacobian between Cartesian and $(\rho, \vartheta, \varphi)$ coordinates, and the parallel wavenumber is given by
\begin{equation}\label{kparallel}
\kp (\rho, \theta) = \kappa_\pi \, (m_1+m_2+\kappa), 
\qquad \kappa_\pi = \sin \Theta / (2\,N_\theta), 
\qquad \kappa = 2\,n\, N_\theta \cot\Theta / R
\end{equation}
where $\Theta$ is the angle between $\boldsymbol{B_{\rm 0}}$ and the toroidal direction and $N_\theta$ the poloidal metric coefficient. With customary poloidal and toroidal angles $\theta$ and $\varphi$, $\kp$ depends on the poloidal angle.
Its expression is symmetric in the field and test function mode indices, a key feature of the tokamak dielectric response theory developed in the Hamiltonian formalism \cite{Gambier1985} as well as in the guiding centre approach of \cite{Lamalle1997}. Other systems of angles may be used, in particular the `constant-$\kp$' (CKP) coordinate system of \cite{Lamalle2006} for which equation (\ref{kparallel}) is replaced by
\begin{equation}\label{cokp}
\kp (\rho) = \kappa_\pi \, (m_1+m_2+\kappa), 
\qquad \kappa_\pi = 1 / (2\,H(\rho)), 
\qquad \kappa = 2\,n\, q(\rho)
\end{equation}
($q$: safety factor, $2\pi H$: field line length per poloidal revolution.) 
The parameters $\kpi$ and $\kappa$ play an important role in the following developments.
%---------------------------------------------------------------------
\section{INTEGRAL KERNEL REPRESENTATION OF THE DIELECTRIC RESPONSE}
Performing inverse poloidal Fourier transforms in (\ref{CTM323bis}) the dielectric response (\ref{CTM323}) can be recast in the form of a poloidal integral, involving the local values of the associated electric field and test function toroidal modes at positions separated by poloidal angle $2\chi$:
\begin{equation}\label{kernelcontrib}
	\W_{\boldsymbol{FE}\beta}^{(n)}
	= \int \, \rmd \rho \; \sum_{p \, = - 1}^{ + 1} \; \sum_{L \, = - 1}^{ + 1} \;
	\int_{-\pi}^{\pi}\int_{-\pi}^{\pi} \, 
	F_{\Ell}^\ast(\rho,\theta\!+\!\chi) \; \K_{\Ell \Ell}^{p} (\theta,\chi) \; E_{\Ell}(\rho,\theta\!-\!\chi)
	\;  \rmd \chi   \,  \rmd \theta
\end{equation}
This representation of the dielectric response involves the integral kernel 
\begin{equation}\label{eq2}
\K_{\Ell \Ell}^{p} (\theta,\chi)= - {\rmi \pi \, RJ} \, \delta_{L,p} \; 2^{\alpha/2} \;
	\frac{\ep \, \omps}{\abs{\kpi} \vT} \;\; \Xi_\alpha(\chi,\kappa,\xi_p)
\qquad (\alpha=2 \delta_{L,0})
\end{equation}
in which the $\rho$ dependence is implicit, and the relevant resonance parameter is
$\xi_p = (\omega-p\omc)/\abs{\kpi\vT}$ 
based on $\kpi$, the elementary poloidal contribution to $\kp$. The coefficients $\xi_p, R, J, \kpi$ and $\kappa$ are all evaluated at $\theta$. New functions of three variables appear, to which we refer as  `kernel dispersion functions' (KDFs):
\begin{equation} \label{eq1}
	\Xi_\alpha(\chi,\kappa,\xi) = \frac{1}{2\pi}\,\sum\limits_{M=-\infty}^{+\infty}  \, \frac{\exp{(\rmi M \chi)}}{\abs{M+\kappa}} \; I_\alpha \left(\frac{\xi}{\abs{M+\kappa}}\right)
\qquad   (\alpha=0, 1, 2)
\end{equation}
The function of index $\alpha\!=\!0$ is relevant to the cyclotron interactions (and to TTMP in the FLR theory); $\alpha\!=\!2$ applies to the Landau-\v{C}erenkov interaction. In the FLR theory, $\alpha\!=\!1$ applies e.g. to the mixed Landau-TTMP contributions. 

Since the Galerkin form (\ref{Galer}) yields Poynting's theorem for $\boldsymbol{F}=\boldsymbol{E}$, the complex HF power absorbed by particle species $\beta$ is given by $P_\beta^{(n)}+\rmi Q_\beta^{(n)}= \W_{\boldsymbol{EE}\beta}^{(n)}$. It is easily verified that the active power $P_\beta^{(n)}$ (resp. reactive power $Q_\beta^{(n)}$) exclusively depends on the imaginary (resp. real) parts of the plasma dispersion functions $I_\alpha$ embedded in the kernel.
%---------------------------------------------------------------------
\section{PROPERTIES OF THE KDFs}

\begin{figure}	\label{ReImXi0vschi_xim2to2by1_ka0p25}
	\centerline{
		\includegraphics*[width=200pt]{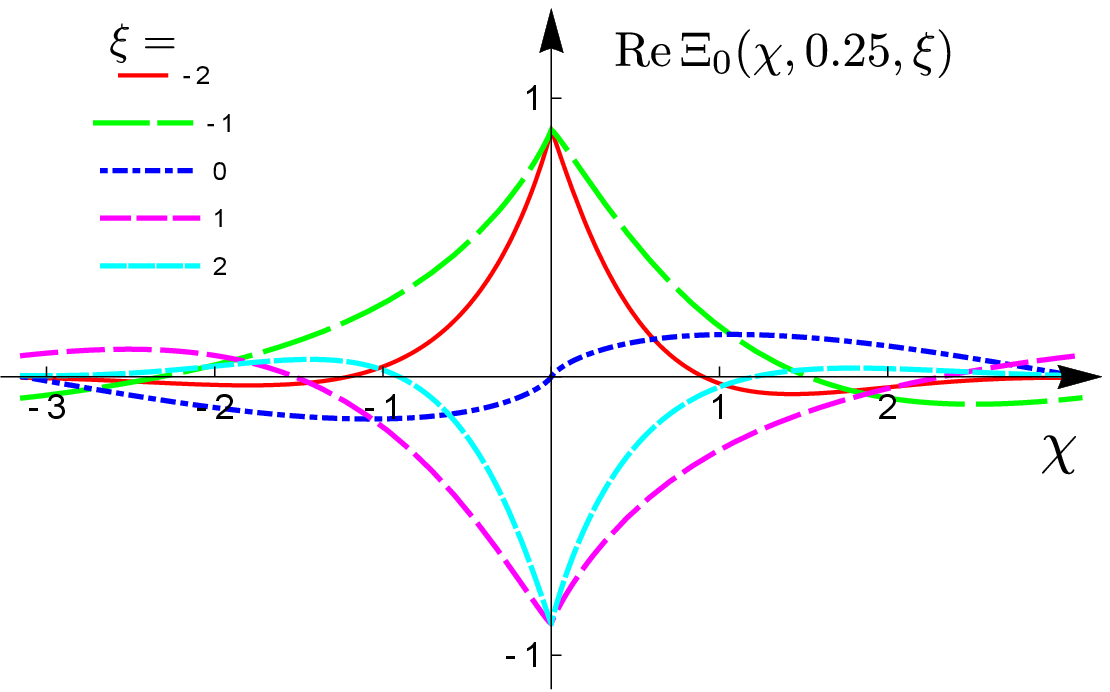}
		\\[10pt]
		\includegraphics*[width=200pt]{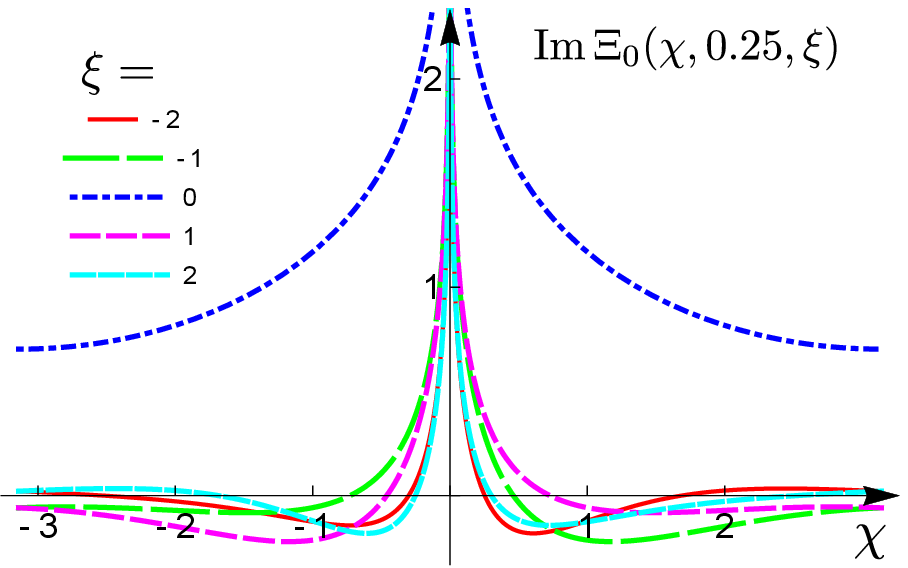}
	}
	\caption{
		Real and imaginary parts of the KDF $\Xi_0(\chi,0.25,\xi)$ vs $\chi$, for a few values of $\xi$.
	}
\end{figure}

The KDFs require careful mathematical treatment since, when $\alpha=0$ or $1$, the series (\ref{eq1}) are conditionally convergent for $\chi\ne 0$ and divergent at $\chi=0$ . Expansions in $\xi$ and $\chi$ series \cite{Lamalle2019a} reveal the logarithmic behaviour with respect to $\chi$ and an integrable singularity of $\Xi_0$ and  $\Xi_1$ at $\chi=0$, see Fig. \ref{ReImXi0vschi_xim2to2by1_ka0p25}; in constrast the $\Xi_2$ series is absolutely convergent and finite at $\chi=0$.

Observing that $\abs{\kp}$ has a minimum over the poloidal spectrum given by
\begin{equation}\label{kpmin}
\abs{\kp}_{\rm min} = \abs{\kappa_\pi} \, \min{(\khat,1-\khat)}, 
\qquad {\rm where} \qquad \khat := \kappa \!\!\! \mod \! 1
\end{equation}
one sees that for each $n\ne0$ the tokamak equilibrium contains surfaces of integer  $\kappa$, where the poloidal spectrum includes a vanishing $\kp$. In CKP coordinates these are simply specific rational-$q$ surfaces. The KDF $\, \Xi_0(\chi,\kappa,\xi_p)$ has a pole at each intersection of the resonance layer $\xi_p=0$ with one of these surfaces, to be handled with the usual causality prescription ${\rm Im} \,\xi_p>0$. (In the degenerate case $n=0$ the singular behaviour occurs all along $\xi_p=0$.)

The following symmetry, $2\pi$-periodicity in $\chi$, and quasi-periodicity in $\kappa$, readily result from the definition (\ref{eq1}): 
\begin{equation} \label{symmetries}
\Xi_\alpha(-\chi,-\kappa,\xi) = \Xi_\alpha(\chi,\kappa,\xi),
\qquad
\Xi_\alpha(\chi+2\pi,\kappa,\xi) = \Xi_\alpha(\chi,\kappa,\xi), 
\qquad
\Xi_\alpha(\chi,\kappa+1,\xi) = \exp(-\rmi \chi) \; \Xi_\alpha(\chi,\kappa,\xi)
\end{equation}
The very interesting quasi-periodicity property was wholly unexpected: it reveals the KDF dependence on the toroidal mode index $n$  embedded in parameter $\kappa$, see (\ref{kparallel})-(\ref{cokp}). It shows that knowledge of the $\Xi_\alpha$ over the $\kappa$ interval $[0,1[$ suffices to evaluate them for arbitrary $\kappa$. Suitable tabulation of the functions should very significantly accelerate computation of the dielectric response in 3D simulations, since the contributions of all toroidal modes can be obtained from a common master table. Taking all symmetries into account the required range of numerical tabulations in ($\chi$, $\kappa$) reduces to either $[0,\pi]\times [0,1[$ or to $]-\pi,\pi]\times [0,1/2]$.
Furthermore tabulations of $\Xi_\alpha$ versus $\xi$ can be limited to the positive range, because the parity properties of the $I_\alpha(\xi)$ ensure that $\Xi_\alpha(-\chi,\kappa,-\xi) = (-)^{\alpha+1}\, \Xi_\alpha^\star(\chi,\kappa,\xi)$ for $\xi$ real.

From a phase integral representation of the standard plasma dispersion function \cite{Stix1992} we have obtained an appealing representation of $\Xi_0$ and $\Xi_2$ in terms of the Jacobian theta function $\vartheta_3(v \, \vert \, \tau)$: in the notation of \cite{EMO1953_II},
\begin{eqnarray} \label{Xi0and2afterjacobi4} 
\left\{\begin{array}{c} \Xi_0(\chi,\kappa,\xi) \\ \Xi_2(\chi,\kappa,\xi) \end{array}\right\} = 
\rmi\frac{\rme^{-\rmi\kappa\chi}}{2\sqrt{\pi}} \int\limits_0^{+\infty}
\exp{\left[ 2\rmi \frac{\xi}{\sqrt{x}} - \frac{\chi^2}{4}x \right]} \; 
\vartheta_3 \! 
\left( \kappa-\rmi \frac{\chi}{2}x\, \vert  \,\rmi \pi x \right)
\left\{\begin{array}{c} 1 \\ -\rmi \xi /\sqrt{x} \end{array}\right\} \frac{\rmd x}{x}
\end{eqnarray}
The properties (\ref{symmetries}) are now seen to ensue from those of the theta function.
Using the method of steepest descent we have derived asymptotic expansions from (\ref{Xi0and2afterjacobi4}), valid for real independent variables and for large values of the product $\abs{\chi\xi}$: introducing $\; \Lambda = (1-\rmi \sigma \sqrt{3}) \, \abs{\chi\xi/2} ^{2/3} /2$, 
with $\sigma= \,\rm{sign}\,\xi$ (the sign of the resonance parameter) the result is
\begin{eqnarray} \label{ahahae3}  
%\left\{\!\begin{array}{c} \Xi_0 \\ \Xi_2 \end{array}\!\right\}
\left\{\begin{array}{c} \Xi_0(\chi,\kappa,\xi) \\ \Xi_2(\chi,\kappa,\xi) \end{array}\right\}
\sim \rmi \, \rme^{-\rmi\kappa\chi} \, \Lambda^{\mp 1/2} \, 
\exp{(-3 \Lambda}) \, / \sqrt{3},
\qquad \qquad \qquad 
\, \abs{\chi\xi}\to +\infty, \; 0<\abs{\chi}<\pi
\end{eqnarray}
(Additional terms are mandatory when $\chi$ approaches $\pm\pi$ where the expansions have Stokes discontinuities.) This very useful result precisely determines the rapid exponential decrease of the KDFs and of the integral kernel (\ref{eq2}): as the average poloidal position $\theta$ recedes from cyclotron resonance, i.e. for large $\abs{\xi_p}$ in (\ref{eq2}), the $\chi$ range of significant coupling between field and test function becomes more and more narrow.

%Plot of $\Xi_0$ vs $\chi$:

%Plot of $\Xi_0$ vs $\xi$:
\begin{figure} \label{ReImXi0vsxi_chi0p3_ka0to1by0p25}
	\centerline{
		\includegraphics*[width=200pt]{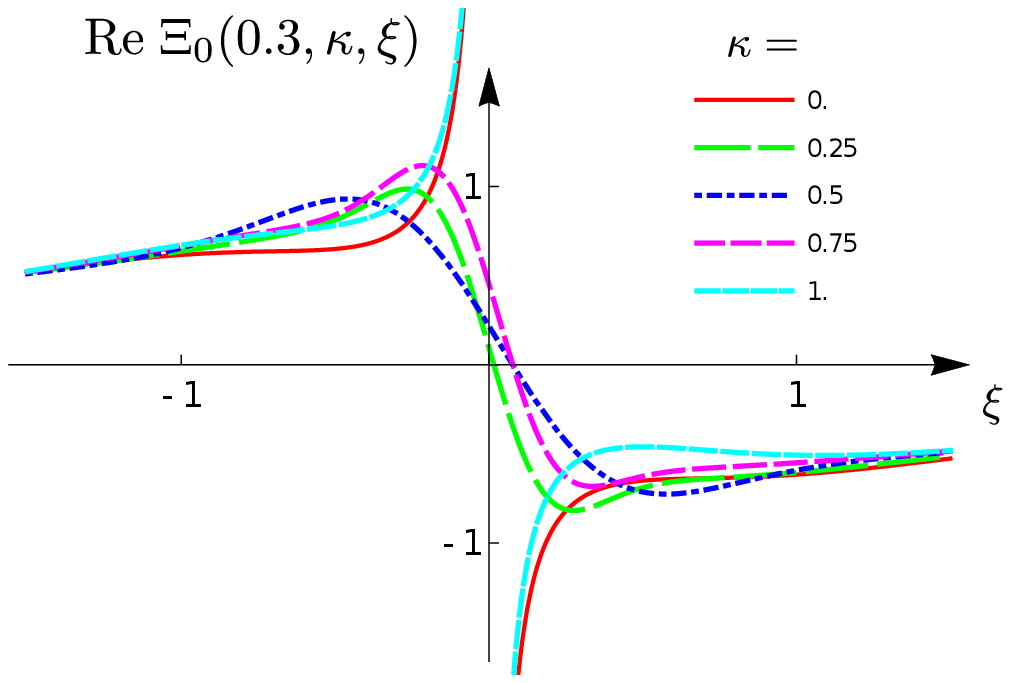} \\[10pt]	
		\includegraphics*[width=200pt]{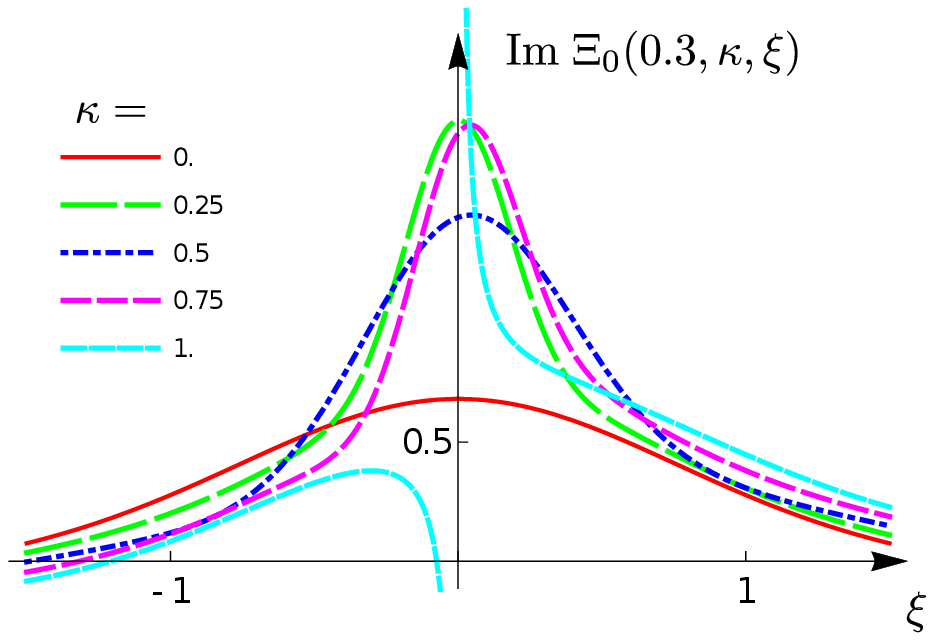}}
	\caption{
		Real and imaginary parts of the KDF $\Xi_0(0.3,\kappa,\xi)$ vs $\xi$, for a few values of $\kappa$.
	}
\end{figure}

\begin{figure} \label{ReImXi0vsxi_chi0p5_ka0p5_ser_ae}
% Plot of series and asymptotic expansion
% Plot of Re $\Xi_0$ vs $\xi$ and $\kappa$
	\centerline{
	\includegraphics[width=230pt]{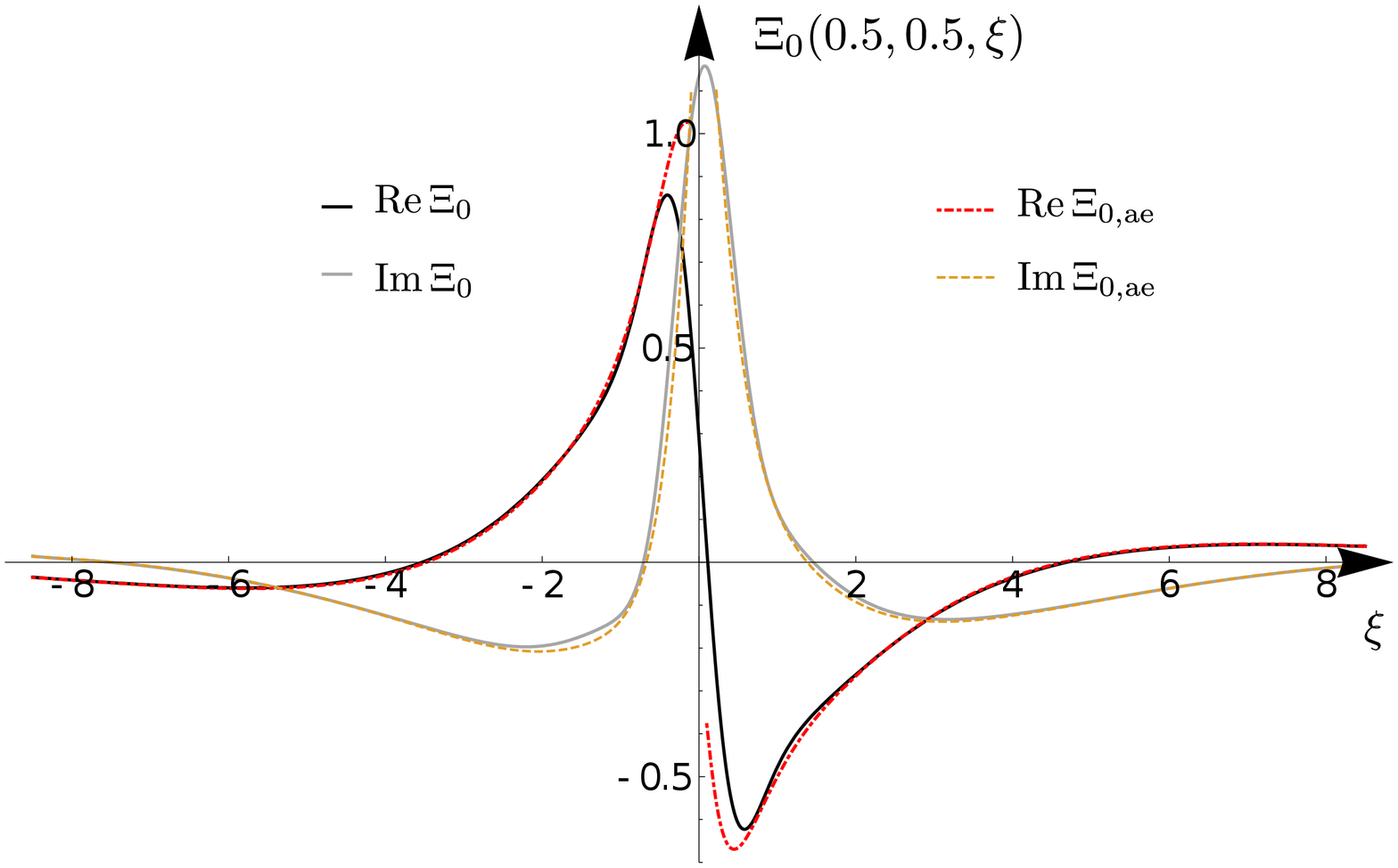}
	\\[10pt]	
	\includegraphics*[width=200pt]{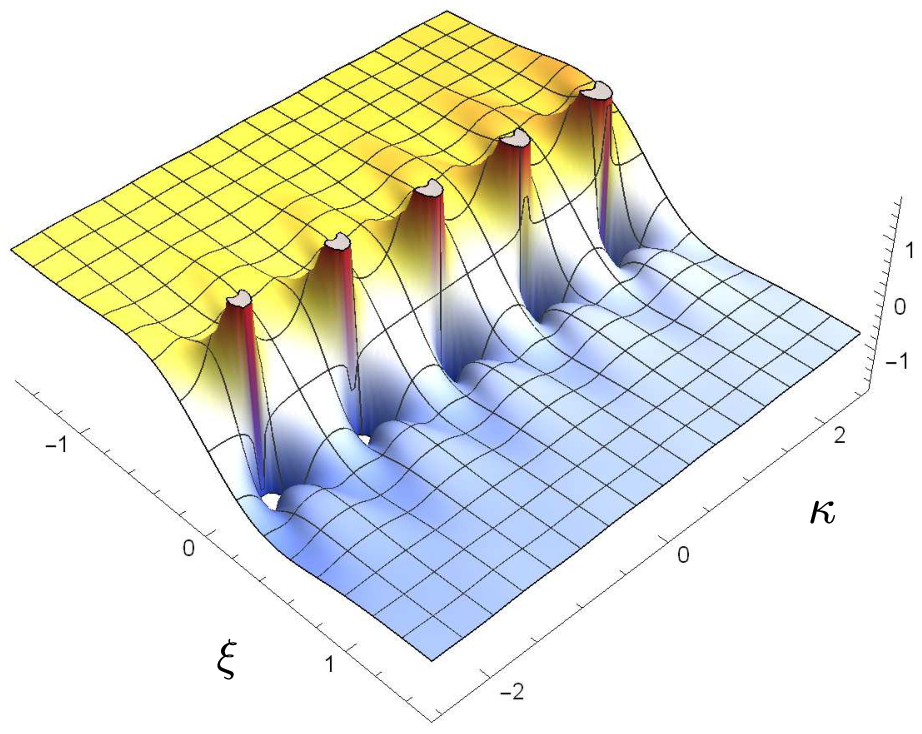} 
	\\[10pt]	
	\includegraphics*[width=15pt]{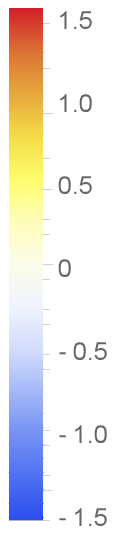}
	}
	\caption{\textbf{Left:} real and imaginary parts of $\Xi_0(0.5,0.5,\xi)$ vs $\xi$; continuous / dashed lines: resp. series / asymptotic expansions. 
	\quad \textbf{Right:} real part of $\Xi_0(0.05,\kappa,\xi)$ vs $\xi$ and $\kappa$. }
\end{figure}

Figure \ref{ReImXi0vschi_xim2to2by1_ka0p25} shows the dependence of $\Xi_0$ upon angle $\chi$, with logarithmic singularity at $\chi\!=\!0$ and asymptotic exponential decay (\ref{ahahae3}) for $\xi\ne 0$. 
Figure \ref{ReImXi0vsxi_chi0p3_ka0to1by0p25} shows the behaviour vs $\xi$ for fixed $\chi$ and $\kappa$. Note the singular resonant behaviour for $\kappa=0$ and $1$. 
The left part of Fig. \ref{ReImXi0vsxi_chi0p5_ka0p5_ser_ae} illustrates the fair agreement between series and asymptotic expansions, in this case for $\abs{\chi\xi} \gtrsim 1$ already. The right part of Fig. \ref{ReImXi0vsxi_chi0p5_ka0p5_ser_ae} shows the combined dependence on $\xi$ and $\kappa$ near cyclotron resonance. 
The dominant variation of $\xi$ is along the tokamak major radius; the one of $\kappa$ mainly manifests itself along the resonance layer. There is a singularity at every intersection between the latter and the surfaces of integer $\kappa$.

%---------------------------------------------------------------------
\section{CONCLUSIONS AND PROSPECTS}
The main results of this work are the kernels and KDFs of Equation \ref{eq2} and \ref{eq1}; the identification of their singularities and asymptotic behaviour allows an efficient numerical treatment; their rapid evaluation by minimal tabulations is straightforward.
The next steps should be implementation in a 2D finite element code, study of wave patterns near cyclotron layers with this new tool, and applications to ICRH scenario simulations. The approach presented here generalizes to all orders in FLR  with the same kernel dispersion functions \cite{Lamalle2019a}; extension to first order in the drift approximation is also feasible. Two-dimensional (poloidal-toroidal) integral kernels can be constructed using the same methods, for applications to massive 3D finite element simulations of wave propagation and absorption in tokamaks.

% References
\bibliographystyle{aipnum-cp}

\bibliography{RFPPC2019_Lamalle}

%Control: year  (1) truncated
\begin{thebibliography}{8}%
\makeatletter
\providecommand \@ifxundefined [1]{%
 \@ifx{#1\undefined}
}%
\providecommand \@ifnum [1]{%
 \ifnum #1\expandafter \@firstoftwo
 \else \expandafter \@secondoftwo
 \fi
}%
\providecommand \@ifx [1]{%
 \ifx #1\expandafter \@firstoftwo
 \else \expandafter \@secondoftwo
 \fi
}%
\providecommand \natexlab [1]{#1}%
\providecommand \enquote  [1]{``#1''}%
\providecommand \bibnamefont  [1]{#1}%
\providecommand \bibfnamefont [1]{#1}%
\providecommand \citenamefont [1]{#1}%
\providecommand \href@noop [0]{\@secondoftwo}%
\providecommand \href [0]{\begingroup \@sanitize@url \@href}%
\providecommand \@href[1]{\@@startlink{#1}\@@href}%
\providecommand \@@href[1]{\endgroup#1\@@endlink}%
\providecommand \@sanitize@url [0]{\catcode `\$12\catcode `\&12\catcode
  `\#12\catcode `\^12\catcode `\_12\catcode `\%12\relax}%
\providecommand \@@startlink[1]{}%
\providecommand \@@endlink[0]{}%
\providecommand \url  [0]{\begingroup\@sanitize@url \@url }%
\providecommand \@url [1]{\endgroup\@href {#1}{\urlprefix }}%
\providecommand \urlprefix  [0]{URL }%
\providecommand \Eprint [0]{\href }%
\providecommand \doibase [0]{http://dx.doi.org/}%
\providecommand \selectlanguage [0]{\@gobble}%
\providecommand \bibinfo  [0]{\@secondoftwo}%
\providecommand \bibfield  [0]{\@secondoftwo}%
\providecommand \translation [1]{[#1]}%
\providecommand \BibitemOpen [0]{}%
\providecommand \bibitemStop [0]{}%
\providecommand \bibitemNoStop [0]{.\EOS\space}%
\providecommand \EOS [0]{\spacefactor3000\relax}%
\providecommand \BibitemShut  [1]{\csname bibitem#1\endcsname}%
\let\auto@bib@innerbib\@empty
%</preamble>
\bibitem [{\citenamefont {Gambier}\ and\ \citenamefont
  {Samain}(1985)}]{Gambier1985}%
  \BibitemOpen
  \bibfield  {author} {\bibinfo {author} {\bibfnamefont {D.~J.}\ \bibnamefont
  {Gambier}}\ and\ \bibinfo {author} {\bibfnamefont {A.}~\bibnamefont
  {Samain}},\ }\href@noop {} {\bibfield  {journal} {\bibinfo  {journal} {Nucl.
  Fusion}\ }\textbf {\bibinfo {volume} {25}},\ p.\ \bibinfo {pages} {283}
  (\bibinfo {year} {1985})}\BibitemShut {NoStop}%
\bibitem [{\citenamefont {Brambilla}\ and\ \citenamefont
  {Kruecken}(1988)}]{Brambilla1988}%
  \BibitemOpen
  \bibfield  {author} {\bibinfo {author} {\bibfnamefont {M.}~\bibnamefont
  {Brambilla}}\ and\ \bibinfo {author} {\bibfnamefont {T.}~\bibnamefont
  {Kruecken}},\ }\href@noop {} {\bibfield  {journal} {\bibinfo  {journal}
  {Nucl. Fusion}\ }\textbf {\bibinfo {volume} {28}},\ p.\ \bibinfo {pages}
  {1813} (\bibinfo {year} {1988})}\BibitemShut {NoStop}%
\bibitem [{\citenamefont {Lamalle}(1997)}]{Lamalle1997}%
  \BibitemOpen
  \bibfield  {author} {\bibinfo {author} {\bibfnamefont {P.~U.}\ \bibnamefont
  {Lamalle}},\ }\href@noop {} {\bibfield  {journal} {\bibinfo  {journal}
  {Plasma Phys. Control. Fusion}\ }\textbf {\bibinfo {volume} {39}},\ \unskip\
  \bibinfo {pages} {1409--1460} (\bibinfo {year} {1997})}\BibitemShut {NoStop}%
\bibitem [{\citenamefont {Lamalle}(2019)}]{Lamalle2019a}%
  \BibitemOpen
  \bibfield  {author} {\bibinfo {author} {\bibfnamefont {P.~U.}\ \bibnamefont
  {Lamalle}},\ }\href@noop {} {\bibfield  {journal} {\bibinfo  {journal}
  {Plasma Phys. Control. Fusion (to be submitted)}\ } (\bibinfo {year}
  {2019})}\BibitemShut {NoStop}%
\bibitem [{\citenamefont {Lamalle}(1998)}]{Lamalle1998}%
  \BibitemOpen
  \bibfield  {author} {\bibinfo {author} {\bibfnamefont {P.~U.}\ \bibnamefont
  {Lamalle}},\ }\href@noop {} {\bibfield  {journal} {\bibinfo  {journal}
  {Plasma Phys. Control. Fusion}\ }\textbf {\bibinfo {volume} {40}},\ \unskip\
  \bibinfo {pages} {465--479} (\bibinfo {year} {1998})}\BibitemShut {NoStop}%
\bibitem [{\citenamefont {Lamalle}(2006)}]{Lamalle2006}%
  \BibitemOpen
  \bibfield  {author} {\bibinfo {author} {\bibfnamefont {P.~U.}\ \bibnamefont
  {Lamalle}},\ }\href@noop {} {\bibfield  {journal} {\bibinfo  {journal}
  {Plasma Phys. Control. Fusion}\ }\textbf {\bibinfo {volume} {48}},\ \unskip\
  \bibinfo {pages} {433--477} (\bibinfo {year} {2006})}\BibitemShut {NoStop}%
\bibitem [{\citenamefont {Stix}(1992)}]{Stix1992}%
  \BibitemOpen
  \bibfield  {author} {\bibinfo {author} {\bibfnamefont {T.~H.}\ \bibnamefont
  {Stix}},\ }\href@noop {} {\emph {\bibinfo {title} {Waves in {P}lasmas}}}\
  (\bibinfo  {publisher} {American Institute of Physics},\ \bibinfo {year}
  {1992})\BibitemShut {NoStop}%
\bibitem [{\citenamefont {Erdélyi}\ \emph {et~al.}(1953)\citenamefont
  {Erdélyi}, \citenamefont {Magnus}, \citenamefont {Oberhettinger},\ and\
  \citenamefont {Tricomi}}]{EMO1953_II}%
  \BibitemOpen
  \bibfield  {author} {\bibinfo {author} {\bibfnamefont {A.}~\bibnamefont
  {Erdélyi}}, \bibinfo {author} {\bibfnamefont {W.}~\bibnamefont {Magnus}},
  \bibinfo {author} {\bibfnamefont {F.}~\bibnamefont {Oberhettinger}}, \ and\
  \bibinfo {author} {\bibfnamefont {F.}~\bibnamefont {Tricomi}},\ }\href@noop
  {} {\emph {\bibinfo {title} {Higher {T}ranscendental {F}unctions}}},\
  Vol.~\bibinfo {volume} {II}\ (\bibinfo  {publisher} {McGraw-Hill},\ \bibinfo
  {year} {1953})\BibitemShut {NoStop}%
\end{thebibliography}%

\end{document}